\documentclass[aps,prl,twocolumn,superscriptaddress,showpacs,preprintnumbers,amsmath,amssymb]{revtex4-1}
\usepackage[utf8]{inputenc}
\usepackage{graphicx}
\usepackage{dcolumn}
\usepackage{notes2bib}
\usepackage{graphicx}
\usepackage{amsmath}
\usepackage{latexsym}
\usepackage{amssymb}
\usepackage{color}
 \usepackage{orcidlink}

\def\CP{C\!P}
\def\Bbar{\overline{B}{}^0}
\def\Dbar{\overline{D}{}^0}

\begin{document}

\title{
\boldmath{Measurements of the branching fraction, polarization, and $\CP$ asymmetry for the decay
$B^0\rightarrow \omega \omega$}}
\noaffiliation
\author{Y.~Guan\,\orcidlink{0000-0002-5541-2278}} 
\author{A.~J.~Schwartz\,\orcidlink{0000-0002-7310-1983}} 
\author{K.~Kinoshita\,\orcidlink{0000-0001-7175-4182}} 
  \author{I.~Adachi\,\orcidlink{0000-0003-2287-0173}} 
  \author{H.~Aihara\,\orcidlink{0000-0002-1907-5964}} 
  \author{S.~Al~Said\,\orcidlink{0000-0002-4895-3869}} 
  \author{D.~M.~Asner\,\orcidlink{0000-0002-1586-5790}} 
  \author{H.~Atmacan\,\orcidlink{0000-0003-2435-501X}} 
  \author{R.~Ayad\,\orcidlink{0000-0003-3466-9290}} 
  \author{S.~Bahinipati\,\orcidlink{0000-0002-3744-5332}} 
% ...
\collaboration{The Belle Collaboration}

\date{\today}

\begin{abstract}
We present a measurement of $B^{0} \rightarrow \omega \omega$, a charmless decay into two vector mesons, 
using 772 $\times 10^6$ $B\overline{B}$ pairs collected with the Belle detector at the KEKB $e^+e^-$ collider.
The decay is observed with a significance of 7.9 standard deviations.
We measure a branching fraction $\mathcal{B} = (1.53 \pm 0.29 \pm 0.17) \times 10^{-6}$,
a fraction of longitudinal polarization $f_L = 0.87 \pm 0.13 \pm 0.13$,
and a time-integrated $\CP$ asymmetry $A_{\CP}$ = $-0.44 \pm 0.43 \pm 0.11$, where 
the first uncertainties listed are statistical and the second are systematic. This is 
the first observation of $B^{0} \rightarrow \omega \omega$, and the first 
measurements of $f_L$ and $A_{\CP}$ for this decay.
\end{abstract}

\maketitle

In the Standard Model (SM), the decay $\Bbar\rightarrow\omega\omega$ proceeds via a $b\rightarrow u$ spectator amplitude and a 
$b\rightarrow d$ loop (``penguin'') amplitude{~\footnote{Charge-conjugate modes are implicitly included throughout this paper, unless noted otherwise}}.
Interference between these two amplitudes, 
which have different weak and strong phases, 
could give rise to direct $\CP$ violation. This
$\CP$ asymmetry can help determine the internal angle
(or phase difference) $\phi^{}_2 \equiv {\rm {arg}}[-(V^*_{tb}V_{td})/(V^*_{ub}V_{ud})]$ of the Cabibbo-Kobayashi-Maskawa (CKM) Unitarity triangle~\cite{Cabibbo:1963yz, Kobayashi:1973fv,Atwood:2001pf}. 
Measuring $\phi^{}_2$ with high precision tests the 
unitarity of the CKM matrix; if the matrix were found to be
non-unitary, that would imply physics beyond the SM such as 
an additional flavor generation.
In addition, the fraction of longitudinal polarization ($f^{}_L$) for this vector-vector
($VV$) final state can also be affected by physics beyond the SM. The value of $f^{}_L$
measured in the $VV$ decay $B\rightarrow \phi K^*$ is surprisingly 
small~\cite{Belle:2003ike,BaBar:2003zor,BaBar:2007bpi,BaBar:2008lan,Belle:2013vat};
this triggered much interest in such decays.
Numerous explanations of this anomaly have been proposed, e.g., 
SM processes such as penguin-annihilation amplitudes~\cite{Li:2004mp} 
and also new physics scenarios~\cite{Baek:2005jk,Bao:2008hd}.
The polarization fraction $f_L$ measured for the color-suppressed decay 
$B^0\rightarrow \rho^0\rho^0$~\cite{Belle:2012ayg,LHCb:2015zxm,BaBar:2008xku,Cheng:2009cn,Zou:2015iwa} is also unexplained;
measurement of $f_L$ for another color-suppressed decay such as 
$B^0 \rightarrow \omega \omega$ would provide insight into the 
QCD dynamics giving rise to different polarization states. 
 
For $B\rightarrow VV$ decays, there are three possible polarization states: 
the longitudinal state with amplitude $H_0$, and two transverse states 
with amplitudes $H_+$ and $H_-$. 
The fraction of longitudinal polarization is defined as 
$f_L \equiv |H_0|^2/(|H_0|^2 + |H_+|^2 + |H_-|^2)$. 
In $B^{0} \rightarrow \omega \omega $ decays, $f_L$ is determined by measuring 
the distribution of the helicity angle $\theta$ of the $\omega$ mesons.
This angle is defined for $\omega \rightarrow \pi^+\pi^-\pi^0$ in 
the $\omega$ rest frame as the difference in direction between that 
of the $B^0$ and the normal to the decay plane of the three pions.

The time-integrated $\CP$ asymmetry is defined as
\begin{equation}
 A_{\CP}\ =\ \frac{\Gamma(\Bbar\rightarrow \omega \omega)  - \Gamma(B^{0} \rightarrow \omega \omega )}
             {\Gamma(\Bbar\rightarrow \omega \omega ) + \Gamma(B^{0} \rightarrow \omega \omega )}\,,
\end{equation} 
where $\Gamma$ is the partial decay width. This asymmetry can differ for each of the 
helicity states, $|H_0|^2$, $|H_+|^2$, and $|H_-|^2$. Measuring $A_{\CP}$ requires
identifying the flavor of the decaying $B$, either $B^0$ or $\Bbar$; this is achieved by
tagging the flavor of the other $B$ produced in $e^+e^-\rightarrow\Upsilon(4S)\rightarrow B^0\Bbar$
reactions that recoils against the signal $B\rightarrow\omega\omega$ decay. 

Theory predictions for the $B^0\rightarrow \omega \omega$ 
branching fraction ($\mathcal{B}$) are in the range $(0.5-3)\times 10^{-6}$; predictions for $f_L$ are in the range $(0.6-0.94)$; and $A_{\CP}$ could be as large as $-$70\%~\cite{Kramer:1991xw,Cheng:2009cn,Li:2005hg,Zou:2015iwa}. 
Experimentally, $B^0\rightarrow \omega \omega$ has been searched for at CLEO-II~\cite{CLEO:1998hej} and 
at {\it BABAR}~\cite{BaBar:2013uwj}. The latter experiment found evidence for this decay with a 
significance of $4.4\sigma$.  No measurement of $f_L$ or $A_{\CP}$ has been reported. In this Letter, 
we report the first observation of $B^0\rightarrow \omega \omega$, and the first measurements of $f_L$ and~$A_{\CP}$.

The data were collected with the Belle detector,
which ran at the KEKB~\cite{KEKB} $e^+e^-$  asymmetric-energy collider.
We analyze the full Belle data set, which corresponds to an integrated 
luminosity of 711\,fb$^{-1}$ containing (771.6 $\pm$ 10.6)$\times 10^{6}$ $B\overline{B}$ pairs ($N_{B\bar{B}}$)
recorded at an $e^+e^-$ center-of-mass (c.m.)\ energy corresponding to the $\Upsilon(4S)$ resonance. 
The Belle detector surrounds the beampipe and consists of several components: 
a silicon vertex detector (SVD) to reconstruct decay vertices; a central 
drift chamber (CDC) to reconstruct tracks; an array of aerogel threshold Cherenkov counters (ACC) and
a barrel-like arrangement of time-of-flight scintillation counters (TOF) to provide particle identification; 
and an electromagnetic calorimeter (ECL) consisting of CsI(Tl) crystals to identify electrons and photons.
All these components are located inside a superconducting solenoid coil providing a 1.5 T magnetic 
field. An iron flux-return located outside the coil is instrumented to identify muons and detect $K^0_L$ 
mesons. More details of the detector can be found in Ref.~\cite{belle_detector}.

We use Monte Carlo (MC) simulated events to optimize selection criteria, calculate signal 
reconstruction efficiencies, and identify sources of background~\cite{Zhou:2020ksj}.
MC events are generated using {\sc EvtGen}~\cite{Lange:2001uf} and 
{\sc Pythia}~\cite{Sjostrand:2000wi}, and subsequently processed through 
a detailed detector simulation using {\sc Geant3}~\cite{Brun:1987ma}. 
Final-state radiation from charged particles is included using {\sc Photos}~\cite{PHOTOS}. 
All analysis is performed using the Belle~II software framework~\cite{Gelb:2018agf}. 

Signal $B^0 \rightarrow \omega \omega$ candidates are reconstructed via the decay chain 
$\omega \rightarrow \pi^+\pi^-\pi^0$, $\pi^0\rightarrow\gamma\gamma$. Reconstructed tracks 
are required to originate from near the $e^+e^-$ interaction point (IP), i.e., have an impact 
parameter with respect to the IP of less than 4.0~cm along the $z$ direction (that opposite 
the direction of the positron beam), and of less than 0.5~cm in the transverse ($x$-$y$) plane. 
Tracks are required to have a transverse momentum of greater than 100~MeV/$c$. To identify 
pion candidates, a particle identification (PID) likelihood is calculated based upon energy-loss 
measurements in the CDC, time-of-flight information from the TOF, and light-yield measurements 
from the ACC~\cite{bellePID}. A track is identified as a pion if the ratio 
$\mathcal{L}(\pi)/[\mathcal{L}(K) + \mathcal{L}(\pi)]>0.4$, where $\mathcal{L}(K)$ 
and $\mathcal{L}(\pi)$ are the likelihoods that a track is a kaon or pion, respectively. The efficiency of this requirement is about 97\%.

Photons are reconstructed from electromagnetic clusters in the ECL that do not
have an associated track. Such candidates are required to have an energy greater than 
50~MeV (100~MeV) in the barrel (end-cap) region, to suppress the beam-induced background.
Candidate $\pi^0$'s are reconstructed from photon pairs that have an 
invariant mass satisfying $M_{\gamma\gamma} \in$ [0.118, 0.150]\,GeV/$c^2$; this range corresponds to  2.5\,$\sigma$ in mass resolution.
In subsequent fits, the invariant mass of photon pairs from $\pi^0$ candidates
are constrained to the nominal $\pi^0$ mass~\cite{PDG2022}. To reduce combinatorial background from 
low-energy photons, we require that the $\pi^0$ momentum be greater than 0.25~GeV/$c$ and that the energies 
of photon pairs satisfy $|E_{\gamma_1} - E_{\gamma_2}|/(E_{\gamma_1} + E_{\gamma_2})< 0.9$.

We reconstruct $\omega\rightarrow\pi^+\pi^-\pi^0$ candidates by combining two oppositely charged 
pion candidates with a $\pi^0$ candidate and requiring that the invariant mass satisfy
$M(\pi^+\pi^-\pi^0) \in [0.74, 0.82]$~GeV/$c^2$; this range corresponds to 4.0\,$\sigma$ in mass resolution. We reconstruct $B^0\rightarrow\omega\omega$ candidates
($B^0_{\rm sig}$) from pairs of $\omega$ candidates that are consistent with originating from a common 
vertex, as determined by performing a vertex fit.
The ordering of the two $\omega$'s is chosen randomly for each event, to avoid an artificial asymmetry in the distribution of helicity angles arising from momentum ordering in the reconstruction.
The particles that are not associated with the signal $B^0\rightarrow\omega\omega$ decay are
collectively referred as ``the rest of the event'' (ROE). We reconstruct a decay vertex for the ROE using tracks in the ROE~\cite{Waltenberger:2008zza}.

To suppress background arising from continuum $e^+e^-\rightarrow q\bar{q}~(q\!=\!u, d, s, c)$ 
production, we use a fast boosted decision tree (FBDT) classifier~\cite{Keck:2017gsv} that distinguishes 
topologically jet-like $q\bar{q}$ events from more spherical $B\overline{B}$ events. 
The variables used in the classifier are: 
modified Fox-Wolfram moments~\cite{KSFW};
CLEO ``cones''~\cite{CLEO:1995rok};
the magnitude of the ROE thrust~\cite{Farhi:1977sg};
the cosine of the angle between the thrust axis of $B^0_{\rm sig}$ and the thrust axis of the ROE;
the cosine of the angle between the thrust axis of $B^0_{\rm sig}$ and the beam axis;
the polar angle of the $B^0_{\rm sig}$ momentum in the $e^+e^-$ c.m.~frame;  
the $p$-value of the $B^0_{\rm sig}$ decay vertex fit; and 
the separation in $z$ between the $B^0_{\rm sig}$ decay vertex and 
vertex of the ROE. 
The classifier is trained using MC-simulated signal decays and $q\bar{q}$ background events.
The classifier has a single output variable, $C_{\rm FBDT}$, which ranges from $-1$ for 
unambiguous background-like events to $+1$ for unambiguous signal-like events.
We require that $C_{\rm FBDT} > 0.75$, which rejects approximately 96\% of $q\bar{q}$ background 
while retaining 78\% of signal events. The variable $C_{\rm FBDT}$ is transformed to a variable 
$C' = \log\left[(C_{\rm FBDT} - 0.75)/(1 - C_{\rm FBDT})\right]$, which is well-modeled by a 
simple sum of Gaussian functions.

To identify $B^0_{\rm sig}$ candidates, we use two kinematic variables: 
the beam-energy-constrained mass $M_{\rm bc}$, and the energy difference $\Delta E$,
defined as 
\begin{eqnarray}
M_{\rm bc} & \equiv & \frac{1}{c^2}\sqrt {E^{2}_{\rm beam} - p^{2}_{B}c^2} \\
\Delta E  & \equiv & E_{B} - E_{\rm beam}\,.
\end{eqnarray}
Here, $E_{\rm beam}$ is the beam energy, and $E_{B}$ and $p_{B}$ are the energy 
and momentum, respectively, of the $B^0_{\rm sig}$ candidate. 
All quantities are evaluated in the $e^+e^-$ c.m.~frame. We retain events 
satisfying $M_{\rm bc} \in [5.24, 5.29]$~GeV/$c^2$ and $\Delta E \in [-0.20, 0.20]$~GeV.

Measuring $A_{\CP}$ requires identifying the $B^0$ or $\Bbar$ flavor of $B^0_{\rm sig}$.
The $B^0\Bbar$ pair produced via $e^+e^-\rightarrow \Upsilon(4S)\rightarrow B^0\Bbar$ 
are in a quantum-correlated state in which the $B^0_{\rm sig}$ flavor must be opposite 
that of the accompanying $B$ at the time the first $B$ of the pair decays. 
The flavor of the accompanying $B$ is identified from inclusive properties of the ROE;
the algorithm we use is described in Ref.~\cite{Belle-II:2021zvj}. The algorithm outputs two quantities: the flavor $q$, where $q\!=\!+1\,(-1)$ corresponds to 
$B^0_{\rm sig}$ being $\Bbar\,(B^0)$, and a quality factor $r$ ranging from 0 for no flavor 
discrimination to 1 for unambiguous flavor assignment. For MC-simulated events, 
$r = 1-2w$, where $w$ is the probability of being mis-tagged.
We do not make a requirement on $r$ but rather divide the data into seven 
$r$ bins with divisions $0.0,\,0.10,\,0.25,\,0.50,\,0.625,\,0.75,\,0.875,\,1.0$. 

The fraction of events having multiple candidates is approximately 10\%.
For these events, the average multiplicity is 2.2. We retain a single candidate by first choosing that with the 
smallest value of $\chi^2(\pi^0_1) + \chi^2(\pi^0_2)$, where $\chi^2(\pi^0)$ 
is the goodness-of-fit resulting from the $\pi^0$-mass-constrained fit
of two $\gamma$ candidates. 
If multiple candidates remain after this selection, we choose that with 
the smallest $\chi^2$ resulting from the $B^0\rightarrow\omega\omega$ vertex fit.
According to MC simulation, these criteria select the correct $B^0_{\rm sig}$ candidate 
in 70\% of multiple-candidate events.

After these selections, the dominant source of background is continuum production, 
which does not peak in $M_{\rm bc}$ or $\Delta E$ but partially peaks in
$M(\pi^+\pi^-\pi^0)$ at $m^{}_\omega$ due to 
$e^+e^-\rightarrow q\bar{q}\rightarrow \omega X$ production. 
For $e^+e^-\rightarrow B\overline{B}$ background, we find that most of this 
background does not peak in $M_{\rm bc}$ or $\Delta E$. 
From MC simulation, we find a small background from $B^0\rightarrow\omega\,{b_{1}(1235)^0} (\rightarrow \omega \pi^0)$ decays, for which the branching fraction is unmeasured.
This background peaks in $M_{\rm bc}$ and at negative 
values of $\Delta E$; thus, we model this background separately when fitting for the signal yield.
Other peaking backgrounds such as $B^0\rightarrow\omega K^{(*)0}$, 
$B^0\rightarrow\omega \eta^{(')}$, $B^0\rightarrow\omega\,{a_{1}(1260)^0}$, 
$B^0\rightarrow\omega\,\pi^+\pi^-\pi^0$, and nonresonant
$B^0\rightarrow \pi^+\pi^-\pi^0 \pi^+\pi^-\pi^0$ decays 
are negligible~\footnote{The decays $B^0\rightarrow\omega\,{a_{1}(1260)^0}$, 
$B^0\rightarrow\omega\,\pi^+\pi^-\pi^0$ and $B^0\rightarrow \pi^+\pi^-\pi^0 \pi^+\pi^-\pi^0$ are currently unmeasured. For this study, we assume branching fractions of 1.0 $\times 10^{-5}$, which is conservatively large -- an order of magnitude larger than $\mathcal{B}(B^0\rightarrow \omega \omega)$}.

The branching fraction, $f_L$, and $A_{\CP}$ are determined from an extended unbinned maximum likelihood 
fit to seven observables.  The fitted observables are $M_{\rm bc}$, $\Delta E$, $C'$, the invariant masses of both 
$\omega$'s [denoted $M_1(\pi^+\pi^-\pi^0)$ and $M_2(\pi^+\pi^-\pi^0)$], and the cosine of the helicity angles 
of both $\omega$'s (denoted $\cos\theta_1$ and $\cos\theta_2$). The fit is performed simultaneously for 
$q=\pm 1$ and for each of seven $r$ bins. The likelihood function is given by

\begin{eqnarray}
\hskip-0.20in
{\cal L}\   =  \frac{\ e^{-(\sum_j N_{j)}}}{\prod_k N_k !} \prod_k \left[\prod_{i=1}^{N_{k}}  \left(\sum_j f_{j,k} N_{j} {\cal P}^i_{j,k}\right) \right],
\end{eqnarray}
where $k$ indicates the $r$-bin,  $N_{k}$ denotes the total number of candidates in the $k^{\rm th}$ bin, $N_{j}$ is the event yield for event category 
$j$ ($j$ = signal, \,$q\bar{q}$, non-peaking $B\overline{B}$ backgrounds, and peaking $B\overline{B}$ backgrounds), $f_{j,k}$ is the fraction of candidates in the $k^{\rm th}$ bin for category $j$, and ${\cal P}^i_{j,k}$ is the corresponding probability density function (PDF). The continuum fractions $f_{{\rm back},k}$ are fixed to values obtained from the data sideband $M_{bc} < 5.265$ GeV$/c^2$. The $B\overline{B}$ background fractions $f_{B\bar{B},k}$ are fixed to values obtained from MC simulation.

The PDF for the signal component is:
\begin{eqnarray}
\mathcal{P}^i_{{\rm sig},k} & = & 
 [1 - q^i \Delta w_k + q^i (1 - 2 w_k) (1 - 2\chi_d) A_{\CP}] \times \nonumber \\
 & & \hskip-0.15in {P}_{{\rm sig},k}(M_{\rm bc}^i, \Delta E^i, C'{}^i, M_1^i, M_2^i, \cos\theta_1^i, \cos\theta_2^i)\,,
\end{eqnarray}
where $q^i\!=\!\pm 1$ is the flavor tag of the $i^{\rm th}$ event, 
$w_k$ is the mistag fraction for bin $k$, 
$\Delta w_k$ is the difference in mistag fractions between $B^0$ tags and $\Bbar$ tags, and
$\chi_d = 0.1858 \pm 0.0011$~\cite{PDG2022} is the time-integrated $B^0$-$\Bbar$ mixing parameter.
The fraction of signal events in the $k^{\rm th}$ bin ($f_{{\rm sig}, k}$), along with $w_k$ and $\Delta w_k$ are determined 
from data using a control sample of $B^0 \rightarrow D^- (\rightarrow K^+ \pi^-\pi^-) \pi^+$ 
decays, in which the final state is flavor-specific (and $B^0$-$\Bbar$ mixing is accounted for)~\cite{Belle-II:2021zvj}. The shape of $C'$ depends slightly 
on $r$ and thus is parameterized separately for each $r$-bin $k$. 

For longitudinally and transversely polarized signal decays, separate PDFs are used, as their $\cos\theta_{1,2}$ distributions differ.
Each PDF in turn consists of two parts, one for correctly reconstructed signal (denoted ``true'')
and one for misreconstructed signal (denoted ``MR''):
$P =  (1 - f_{\rm MR})\,{P_{\rm true}} + f_{\rm MR}\,{P_{\rm MR}}$.
The fraction of misreconstructed signal ($f_{\rm MR}$) is fixed from MC simulation; this
value is 14.6\% (17.6\%) for longitudinally (transversely) polarized decays.

For correctly reconstructed signal, 
the $M_{\rm bc}$ distribution is modeled by a Crystal Ball function~\cite{CB}.
The $(\Delta E, M_1, M_2)$ distribution is modeled by a three-dimensional 
histogram that accounts for correlations among these observables, and
$C'$ is modeled by the sum of a Gaussian distribution and a bifurcated Gaussian. The $\cos\theta_{1,2}$ distributions 
are modeled by a histogram from MC simulation.
For mis-reconstructed signal,
$(M_{\rm bc}, \Delta E)$ is modeled by a two-dimensional histogram that accounts for correlations, and
$C'$ is modeled by the sum of two Gaussian functions. 
The variables $M_{1,2}$ and  $\cos\theta_{1,2}$ are modeled by histograms from MC simulation. 
To account for differences between data and MC simulation, 
the PDFs for $M_{\rm bc}$, $\Delta E$, $C^{'}$ and $M_{1,2}$
are adjusted with calibration factors determined from a control 
sample of $B^0\rightarrow\Dbar(\rightarrow K^+\pi^-\pi^0)\omega$ decays.

For continuum background, correlations among observables are negligible. 
The $M_{\rm bc}$ distribution is modeled by a threshold ARGUS~\cite{ARGUS:1990hfq} function,
$\Delta E$ is modeled by a second-order polynomial, and 
$C'$ is modeled by the sum of two Gaussian functions.
The PDFs for $M_{1,2}$, $\cos\theta_{1,2}$, are divided into two parts to account
for true and falsely reconstructed (denoted ``non-$\omega$'') $\omega\rightarrow\pi^+\pi^-\pi^0$ decays:
\begin{eqnarray}
{P}_{q\bar{q}}(M,\cos\theta)\  & = & \ 
f_{\omega}\,{P}_{\omega}(M)\,{P}_{\omega}(\cos\theta) + \nonumber \\
 & & \hskip-0.20in (1-f_{\omega})\,{P}_{{\rm non}\mbox{-}\omega}(M)\,{P}_{{\rm non}\mbox{-}\omega}(\cos\theta)\,.
\end{eqnarray}
The fraction of $q\bar{q}$ background containing true $\omega$ decays ($f_{\omega}$) is floated in the fit.
For these decays, $M_1$ and $M_2$ are modeled by a histogram from MC simulation, and the $\cos\theta_{1,2}$ 
distributions are modeled by polynomials. The PDFs for the non-$\omega$ component are taken 
to be polynomials. All shape parameters except those for $C'$ are floated in the fit; the 
shape for $C'$ is fixed to that from MC simulation. 
The PDFs for $C'$ and $M_{1,2}$ for true $\omega$'s 
are adjusted with small calibration factors determined from the $B^0\rightarrow\Dbar(\rightarrow K^+\pi^-\pi^0)\omega$ control sample.

For non-peaking $B\overline{B}$ background, 
$M_{\rm bc}$ is modeled by a threshold ARGUS function,
$\Delta E$ is modeled by a second-order polynomial,
$C'$ is modeled by the sum of two Gaussian functions, 
and $M_{1,2}$ and $\cos\theta_{1,2}$ are modeled by histograms from MC simulation.
For peaking $B\overline{B}$ background, all PDF shapes are obtained from histograms from MC simulation.

There are a total of 
16 floated parameters in the fit: 
the yields of signal, continuum, peaking $B\overline{B}$, and non-peaking $B\overline{B}$ backgrounds, 
the parameters $f_L$ and $A_{\CP}$, 
and PDF parameters (except that for $C'$) for $q\bar{q}$ background.
We fit directly for the branching fraction (${\cal B}$) using the 
relation between ${\cal B}$ and the signal yields:

\begin{equation}
\begin{aligned}
& N_L = 2 \times N_{B^0\Bbar} \times \mathcal{B}  \times \mathcal{B}_\omega \times f_L \times \varepsilon_L&  \\
& N_T= 2 \times N_{B^0\Bbar} \times \mathcal{B}  \times \mathcal{B}_\omega \times ( 1- f_L ) \times \varepsilon_T.&
\end{aligned}
\end{equation}
where $N_L$ ($N_T$) is the yield of longitudinally (transversely) polarized signal and $N_L + N_T = N^{}_{\rm sig}$, $N_{B^0\Bbar}$ is the number of $B^0\Bbar$ pairs, $\mathcal{B}_\omega = [\mathcal{B}(\omega \rightarrow \pi^+\pi^-\pi^0) \times 
\mathcal{B} (\pi^0 \rightarrow  \gamma\gamma)]^2$, and
$\varepsilon_L$ and $\varepsilon_T$ are the signal reconstruction 
efficiencies. We take $N_{B^0\Bbar}$ to be $N_{B\bar{B}}\times f^{00}$, where $f^{00}$ = 0.484 $\pm$ 0.012 is the fraction of $B^0\Bbar$ production at the $\Upsilon(4S)$~\cite{Belle:2022hka}. The efficiencies $\varepsilon_L$ and $\varepsilon_T$ are obtained from MC simulation as the ratio of the number of events that pass all selection criteria to the total number of simulated events. 
We find $\varepsilon_L = 8.82 \pm 0.02$ (\%) and $\varepsilon_T = 6.54 \pm 0.02$ (\%), respectively, for longitudinally and transversely
polarized signal decays.

\begin{figure}
\begin{center}
\includegraphics[width=0.48\textwidth]{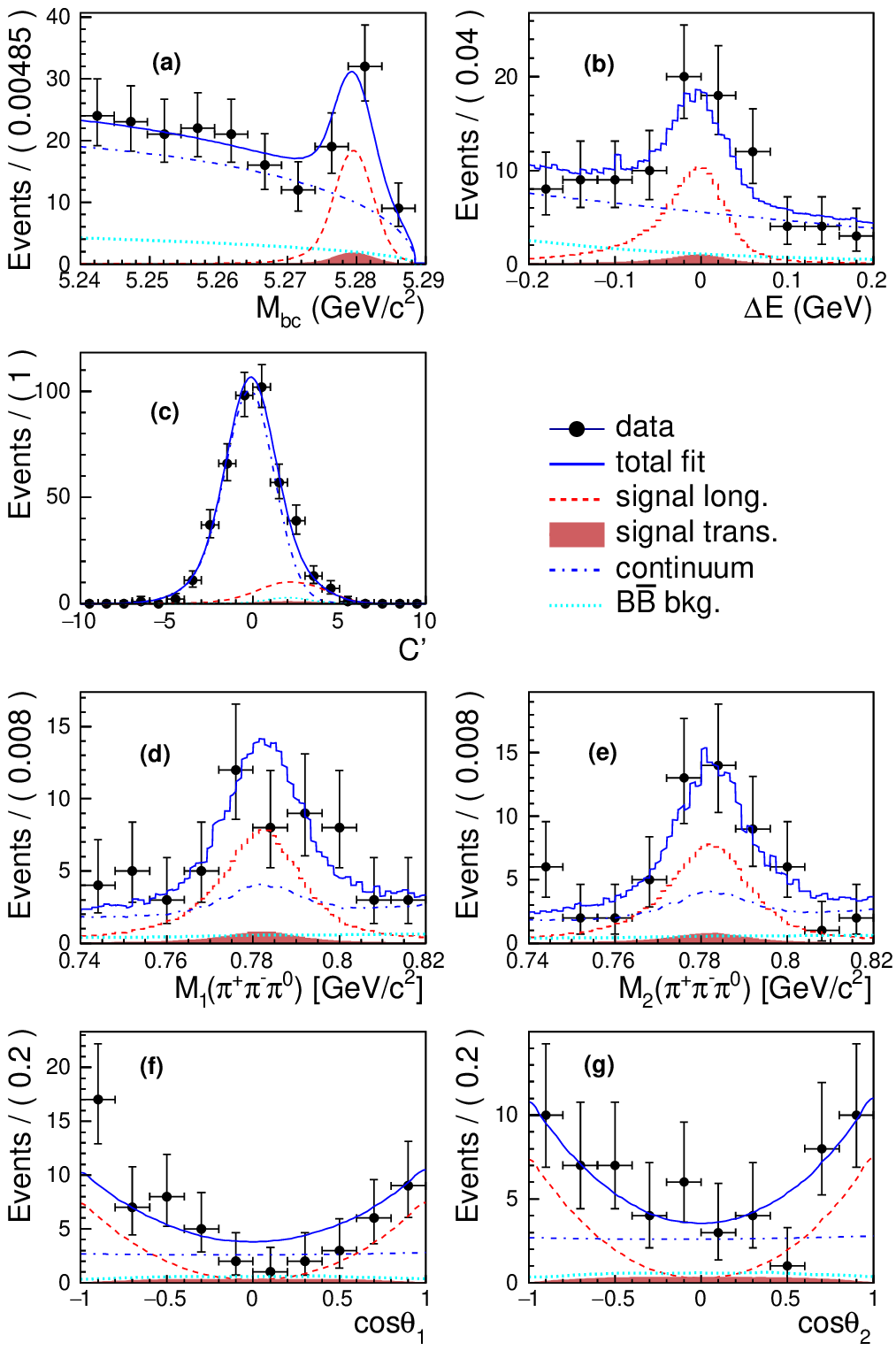}
\caption{Projections of the fit for 
(a)~$M_{\rm bc}$ (b)~$\Delta E $ (c)~$C'$ (d)~$M_1(\pi^+\pi^-\pi^0)$ 
(e)~$M_2(\pi^+\pi^-\pi^0)$ (f)~$\cos\theta_1$, and (g)~$\cos\theta_2$.
Events plotted are in a signal-enhanced region (except for the variable plotted) 
of $M_{\rm bc}\!\in\![5.274, 5.290]$~GeV/$c^2$,  $\Delta E\!\in\![-0.080,0.080]$~GeV, 
and $C'\!\in\![2, 10]$.
The red-dashed line shows longitudinally polarized signal;
the red-shaded area shows transversely polarized signal; 
the blue-dash-dotted line shows the $q\bar{q}$ background;
the cyan-dotted line shows the $B\overline{B}$ background,
and the blue-solid curve shows the overall fit result.}
\label{fig:fit_data_qrsum_enhance}
\end{center}
\end{figure}

The projections of the fit are shown in Fig.~\ref{fig:fit_data_qrsum_enhance}.
We obtain
$N_{\rm sig} = 60.3\pm 10.8$, $f_L =  0.87 \pm 0.13$, and $A_{\CP} = -0.44\pm 0.43$. The significance of the signal
is evaluated  using the difference of the likelihoods for the nominal fit and for a fit with the signal yield set to zero. 
In the later case, there are three fewer degrees of freedom: the signal yield, $f_L$ and $A_{CP}$.
Systematic uncertainties are included in the significance calculation 
by convolving the likelihood function with a Gaussian function whose width is
equal to the total additive systematic uncertainty (see Table~\ref{table:sys}). 
The signal significance including systematic uncertainties corresponds to $7.9\,\sigma$.

The systematic uncertainties 
are summarized in Table~\ref{table:sys}.
The uncertainty due to the reconstruction efficiency has several contributions:
charged track reconstruction (0.35\% per track), 
$\pi^0$ reconstruction (4.0\%~\cite{Belle:2014mfl}), 
PID efficiency (3.5\%), and continuum suppression (2.4\%).
The systematic uncertainty due to continuum suppression is evaluated using the $B^0\rightarrow \Dbar(\rightarrow K^+\pi^-\pi^0)  \omega$ control 
sample:
the requirement on $C_{\rm FBDT}$ is varied and the resulting change in the
efficiency-corrected yield (2.4\%) is assigned as a systematic uncertainty for $\mathcal{B}$.
The uncertainty due to the best-candidate selection is evaluated by randomly choosing a candidate;
the resulting changes in $\mathcal{B}$, $f_{L}$, and $A_{\CP}$ are assigned as systematic uncertainties.
The systematic uncertainty due to calibration factors for PDF shapes is evaluated by varying these
factors by their uncertainties and repeating the fits. The resulting variations in the fit results 
are assigned as systematic uncertainties.
The correlation between $\Delta E$ and $M_1$ and $M_2$ as found from 
MC simulation is accounted for in the fit, but there could be differences 
between the simulation and data. We thus change this correlation by $10\%$ (absolute) and refit the data; the changes in the fit results are assigned as systematic uncertainties.  
The fraction of mis-reconstructed signal is varied by $\pm 30$\% and the changes from the 
nominal results are assigned as systematic uncertainties.
From a large ``toy'' MC study, small potential biases are observed in the fit results. 
We assign these biases as systematic uncertainties. Finally, we include uncertainties 
on ${\cal B}$ arising from $N_{B^0\Bbar}$ (2.8\%) and intermediate branching fractions
${\cal B}(\omega\rightarrow \pi^+\pi^-\pi^0)\times {\cal B}(\pi^0\rightarrow\gamma\gamma)$ (1.6\%)~\cite{PDG2022}.

For $A_{\CP}$, there is systematic uncertainty arising from flavor tagging. We evaluate 
this by varying the flavor-tagging parameters $\varepsilon_k$,  $w_k$, and $\Delta w_k$ by 
their uncertainties; the resulting change in $A_{\CP}$ is taken as a systematic uncertainty.
To account for a possible asymmetry in backgrounds (arising, e.g., from the detector), 
we include an $A_{\CP}^{ q\bar{q}}$ term in the continuum background PDF. We float 
$A_{\CP}^{q\bar{q}}$ in the fit and obtain a value $0.008 \pm 0.014 $, which is 
consistent with zero. The resulting change in the signal $A_{\CP}$ is assigned 
as a systematic uncertainty. The total systematic uncertainties are obtained by 
combining all individual uncertainties in quadrature; the results are
11.4\% for ${\cal B}$, 0.13 for $f_L$, and 0.11 for $A_{\CP}$.

\begin{table}
\renewcommand{\arraystretch}{1.2}
\caption{Systematic uncertainties on ${\cal B}$, $f_L$, and $A_{\CP}$. Those listed in 
the upper part are additive and included in the significance calculation as discussed in the text. Those listed in the lower part are multiplicative.}
\label{table:sys}
\begin{tabular}{lccc}
\hline
\hline
Source  &   ${\cal B}$ (\%)  &   $f_{L}$ & $A_{\CP}$   \\
\hline
Best candidate selection & 3.0  & 0.07  & 0.04  \\
Signal PDF & 7.7  & 0.10 & 0.10 \\
Fit bias & 3.0 &  0.01 &  0.01\\
Background PDF & 0.7 &  0.00 & 0.01 \\
\hline
Tracking efficiency  & 1.4  &  0.00   &  0.00   \\
$\pi^0$ efficiency & 4.0 & 0.00  & 0.00  \\
PID efficiency & 3.5 &  0.00  & 0.00  \\
Continuum suppression  & 2.4 & $-$  & $-$   \\
Flavor mistagging & $-$  &$-$   & 0.02\\
Detection asymmetry  & $-$  &  $-$  & 0.01 \\
$N_{B^0\Bbar}$ & 2.8 & $-$  & $-$  \\
${\cal B}(\omega\rightarrow \pi^+\pi^-\pi^0)\times {\cal B}(\pi^0\rightarrow\gamma\gamma)$ 
& 1.6 & $-$  & $-$ \\
\hline
Total & 11.4 & 0.13 & 0.11 \\
\hline
\hline
\end{tabular}
\end{table}

In summary, we report measurements of the decay $B^{0}\rightarrow\omega\omega$ 
using $772\times 10^6$ $B\overline{B}$ pairs produced at the Belle experiment. The branching 
fraction, fraction of longitudinal polarization, and time-integrated $\CP$ asymmetry 
are measured to be
\begin{eqnarray}
\mathcal{B} & = & (1.53 \pm 0.29 \pm 0.17) \times 10^{-6} \\
f_L         & = & 0.87 \pm 0.13 \pm 0.13 \\
A_{\CP} & = & -0.44 \pm 0.43 \pm 0.11\,, 
\end{eqnarray}
where the first uncertainties are statistical and the second are systematic. 
The $B^{0}\rightarrow\omega\omega$ decay is observed for the first time; the 
significance including systematic uncertainties is ${7.9}\,\sigma$.  Our measurements 
of $f_L$ and $A_{\CP}$ are the first such measurements.
Our results for ${\cal B}$ and $f_L$ are consistent with theoretical estimates, 
while our result for $A_{\CP}$ shows no significant $\CP$ violation.

\begin{acknowledgments}
This work, based on data collected using the Belle detector, which was
operated until June 2010, was supported by 
the Ministry of Education, Culture, Sports, Science, and
Technology (MEXT) of Japan, the Japan Society for the 
Promotion of Science (JSPS), and the Tau-Lepton Physics 
Research Center of Nagoya University; 
the Australian Research Council including grants
DP210101900, 
DP210102831, 
DE220100462, 
LE210100098, 
LE230100085; 
Austrian Federal Ministry of Education, Science and Research (FWF) and
FWF Austrian Science Fund No.~P~31361-N36;
National Key R\&D Program of China under Contract No.~2022YFA1601903,
National Natural Science Foundation of China and research grants
No.~11575017,
No.~11761141009, 
No.~11705209, 
No.~11975076, 
No.~12135005, 
No.~12150004, 
No.~12161141008, 
and
No.~12175041, 
and Shandong Provincial Natural Science Foundation Project ZR2022JQ02;
the Czech Science Foundation Grant No. 22-18469S;
Horizon 2020 ERC Advanced Grant No.~884719 and ERC Starting Grant No.~947006 ``InterLeptons'' (European Union);
the Carl Zeiss Foundation, the Deutsche Forschungsgemeinschaft, the
Excellence Cluster Universe, and the VolkswagenStiftung;
the Department of Atomic Energy (Project Identification No. RTI 4002), the Department of Science and Technology of India,
and the UPES (India) SEED finding programs Nos. UPES/R\&D-SEED-INFRA/17052023/01 and UPES/R\&D-SOE/20062022/06; 
the Istituto Nazionale di Fisica Nucleare of Italy; 
National Research Foundation (NRF) of Korea Grant
Nos.~2016R1\-D1A1B\-02012900, 2018R1\-A2B\-3003643,
2018R1\-A6A1A\-06024970, RS\-2022\-00197659,
2019R1\-I1A3A\-01058933, 2021R1\-A6A1A\-03043957,
2021R1\-F1A\-1060423, 2021R1\-F1A\-1064008, 2022R1\-A2C\-1003993;
Radiation Science Research Institute, Foreign Large-size Research Facility Application Supporting project, the Global Science Experimental Data Hub Center of the Korea Institute of Science and Technology Information and KREONET/GLORIAD;
the Polish Ministry of Science and Higher Education and 
the National Science Center;
the Ministry of Science and Higher Education of the Russian Federation
and the HSE University Basic Research Program, Moscow; 
University of Tabuk research grants
S-1440-0321, S-0256-1438, and S-0280-1439 (Saudi Arabia);
the Slovenian Research Agency Grant Nos. J1-9124 and P1-0135;
Ikerbasque, Basque Foundation for Science, and the State Agency for Research
of the Spanish Ministry of Science and Innovation through Grant No. PID2022-136510NB-C33 (Spain);
the Swiss National Science Foundation; 
the Ministry of Education and the National Science and Technology Council of Taiwan;
and the United States Department of Energy and the National Science Foundation.
These acknowledgements are not to be interpreted as an endorsement of any
statement made by any of our institutes, funding agencies, governments, or
their representatives.
We thank the KEKB group for the excellent operation of the
accelerator; the KEK cryogenics group for the efficient
operation of the solenoid; and the KEK computer group and the Pacific Northwest National
Laboratory (PNNL) Environmental Molecular Sciences Laboratory (EMSL)
computing group for strong computing support; and the National
Institute of Informatics, and Science Information NETwork 6 (SINET6) for
valuable network support. We thank Justin Albert (University of Victoria) for very helpful discussions.
\end{acknowledgments}
\bibliography{omega.bib}
\end{document}